\documentclass{elsart}
\usepackage{graphicx}
\usepackage{amssymb}
\def\sgra{Sgr\,A*}
\def\jpsi{\langle J(0)\rangle_{\Delta\Omega}}
\def\jpsiDO{\jpsi\,\Delta\Omega}

\def\ml{12~TeV}
\begin{document}
\begin{frontmatter}
\title{TeV $\gamma$-radiation from Dark Matter annihilation in the Galactic center}
\author{D. Horns}
\address{Max-Planck-Institut f\"ur Kernphysik,
Postfach 10\,39\,80, D-69029 Heidelberg, Germany}

\begin{abstract}

 Recent observations of the Galactic center with ground based gamma-ray
instruments (CANGAROO-II, VERITAS, H.E.S.S.) have revealed a TeV (10$^{12}$~eV) gamma-ray signal consistent with
the position of \sgra.
  The derived luminosity of the signal above 1 TeV is
a few $10^{34}~\mathrm{erg}\,\mathrm{s}^{-1}$ which is slightly more than e.g.
the gamma-ray luminosity of the Crab nebula and a plausible identification
could be a conventional albeit strong gamma-ray source.  
The observations with the H.E.S.S. system of Cherenkov
telescopes have decreased the solid angle subtended by the error box by more
than a factor of hundred with respect to the previous observations of the
VERITAS and CANGAROO groups.  After
studying the observed energy spectrum and angular distribution of the excess as
seen by the H.E.S.S. experiment, a massive Dark Matter candidate with a minimum
mass of \ml  (90~\% c.l.) and an upper limit on the WIMP density for $r<10$~pc
of $1261~M_\odot\,\mathrm{pc}^{-3} \times \left(\langle \sigma
v\rangle/3\cdot10^{-26}\,\mathrm{cm}^3\mathrm{s}^{-1}\right)^{-1/2}$ is
required to explain the observed flux by an annihilation signal. The angular
distribution of the excess events  is consistent with a cuspy profile with
$\rho(r)\propto r^{-\alpha}$ with $\alpha>1.0$ at a confidence level of 90~\%.
Even though the mass and the cross section of the Dark Matter constituents are
unexpectedly high in the framework of most models of nonbaryonic Dark Matter,
it can not be ruled out.  
\end{abstract}
\begin{keyword}
\PACS 95.35.+d \sep 98.35.Jk \sep 98.35.Gi \sep 95.85.Pw \sep 96.40.Pq

\end{keyword}
\end{frontmatter}

\section{Introduction}

The presence of a Dark Matter halo of the Galaxy is observationally established
from stellar dynamics \cite{2002ApJ...573..597K}.  At
present, the nature of Dark Matter is unfortunately unknown but a number of
viable candidates have been advocated within different theoretical frameworks
mainly motivated by particle physics (for  recent reviews see e.g.
\cite{2000RPPh...63..793B,hep.ph.0404175}) including the widely studied models of
Supersymmetric (SuSy) Dark Matter \cite{1984NuPhB.238..453E}.  On the
cosmological scale, the presence of non-baryonic and predominantly cold Dark
Matter as it had been suggested in \cite{1984Natur.311..517B} is well
established by many observations most notably of the WMAP experiment
\cite{2003ApJS..148..175S}.

 The implications of the possible presence of SuSy Dark Matter in the Galactic
center has been studied in numerous publications including predictions for the
$\gamma$-ray flux expected from annihilation of \textit{weakly interacting
massive particles} - WIMPs (see e.g.  \cite{1998APh.....9..137B,2003MNRAS.345.1313S,2004PhRvD..69c5001H,2004PhRvD..69d3501K,2004PhRvD..69l3501E}). With the advent of the next generation of Cherenkov
telescopes a gamma-ray signal from the direction of the Galactic center has
been established:
	 The recent claims for  gamma-rays from the Galactic center by the CANGAROO
collaboration with a signal $S=9.8~\sigma$ \cite{2004ApJ...606L.115T} and a
possible excess claimed by the VERITAS collaboration on the level of
3.7~$\sigma$ \cite{2004ApJ...608L..97K} have already stimulated first
speculations on the interpretation of the signal as annihilation radiation from
Dark Matter \cite{2004astro.ph..4205H}. Finally, the recently commissioned
H.E.S.S. system of Cherenkov telescopes has detected a signal from the vicinity
of \sgra \cite{Nature}. The different confidence regions of the detections by
the different instruments in comparison with a 21~cm radio map
\cite{1989IAUS..136..243Y} are  indicated in Fig.~1 and the flux measurements
are given in Fig.~2. Note, the EGRET measurements taken from
\cite{1998A&A...335..161M,1999ApJS..123...79H} give the range of possible flux
values for non-resolved objects in the gamma-ray background dominated region
and should be interpreted as upper limits for a contribution of \sgra. The
upper limit indicated in Fig.~2 at 1~GeV is from a dedicated analysis of the
high energy EGRET data \cite{2002astro.ph.10617H} assuming a photon index of
1.5.

At first glance, the differences in the reported TeV flux by the CANGAROO and
H.E.S.S. groups might suggest variability at the lower end of the energy
spectrum thus ruling out an exclusive annihilation origin and hinting at a more
conventional compact gamma-ray source related to \sgra. 
However, given the positional uncertainty of the CANGAROO II and VERITAS
results evident in Fig.~1, it remains possible that in fact more than one
(possibly even variable) 
source may exist in the direction of the Galactic center.  Interestingly, the
observations of H.E.S.S. indicate the presence of even 
further sources in the field
of view (see Fig.~1 of \cite{Nature}). 

Further observations with H.E.S.S. and CANGAROO III in the coming years will
help to clarify the situation which might be consistent with multiple sources
which are not resolvable by single telescope observations.  Instrumentally, the
CANGAROO and H.E.S.S. observations have shown discrepancies in the flux values
for other steady sources: SN1006, PSR B1706-44, Vela, and RXJ1713-304 \cite{Rapporteur,Komin,berg} -  a similar
situation might be present in this case.  

The VERITAS energy flux given in Fig.~2 is converted from the 40~\% Crab flux
as measured with the Whipple telescope, assuming a photon index of 2.2, and
using the Crab flux  as determined by H.E.S.S. to compare the two instruments'
observations.  The error bars are given as the sum of statistical and
systematic uncertainties.  The two measurements are in reasonable agreement and
given the possible positive bias in the flux estimate for a source detected at
the level  of less than 4 standard deviations should be considered to be
consistent within the uncertainties.

The discussion here is focussing on the results obtained with the H.E.S.S.
system of telescopes which achieves a superior angular resolution in comparison
to the previously published results as is evident from
Fig.~1\,\,\footnote{These data were taken during the build-up phase and will
eventually be superseded by the performance of the full telescope array}. The
reconstructed source location is, within the errors, consistent with the
position of \sgra (space angle $\Delta \theta=(18\pm42)$'') and appears to be
consistent with a pointsource for the given angular resolution of $5.8$' (50~\%
containment radius  \cite{Nature}).

 The H.E.S.S. observations (spectrum and angular distribution) are interpreted
in the framework of a Dark Matter annihilation scenario to investigate the
consequences for the WIMP mass, density, and spatial structure and to
understand whether such an interpretation is feasible.

\begin{figure}
\centering
\includegraphics[width=\linewidth]{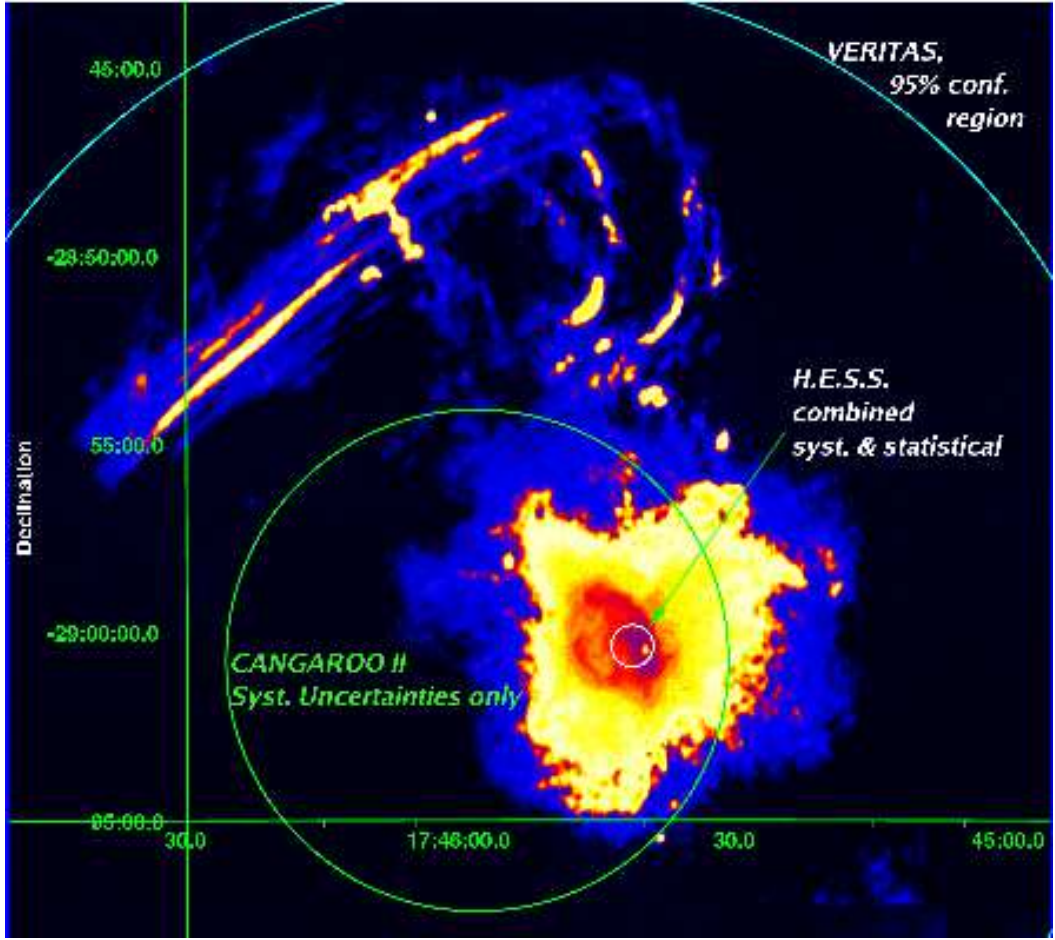}

\caption{Position of the TeV emitting region overlaid on a 21~cm radio map
\cite{1989IAUS..136..243Y} tracing mostly non-thermal (synchrotron) emission
features.  The bright spot inside the H.E.S.S. confidence region (marked as a
white circle) is \sgra whereas the extended ring like feature to the east of it
is \sgra East.  Note the difference in the accuracy of the different
instruments.} 

\end{figure}

\section{WIMP mass}

 In order to quantify the minimum mass required to explain the observed energy
spectrum in terms of an annihilation signal, an assumption on the shape of the
annihilation continuum spectrum is required. In the very generic case of a
self-annihilating massive WIMP, it is assumed that the branching
ratio for $q\bar q$, $W^+W^-$, and $Z^0Z^0$ is quite large. 

Subsequent hadronization leads to the production of charged and neutral mesons
which decay finally into electrons, neutrinos, and gammas. The gamma signal is
described by a fairly hard differential energy spectrum with an exponential
cut-off $d\Phi/dE\propto E^{-1.5}\exp(-E/E_{c})$ with $E_c\approx 0.1 m_\chi$
\cite{1998APh.....9..137B}.  However, the exact shape of the spectrum depends
on details of the branching ratio of other processes. Namely, for
$\chi\chi\rightarrow \tau \bar \tau$, direct radiative losses produce a second
hard component. 

 For the purpose of calculating a lower limit on $m_\chi$ it is assumed that
the branching ratio for $\chi\chi\rightarrow q\bar q$ or $W^+W^-$is close to
one. The minimum mass requirement is somewhat relaxed if $\chi\chi\rightarrow
\tau\bar\tau$ is of importance.

 Fitting a generic $\pi^0$ decay spectrum to the data results in a 90~\% c.l.
lower limit on the mass $m_\chi$ to be larger than \ml.

In most models that attempt to explain non-baryonic Dark Matter in the
Universe, the relic particles are WIMPs produced thermally during the early
phase of cosmological expansion. A very reasonable candidate for this particle
is provided by SuSy models which naturally predict the existence of a stable
(if R-parity is conserved) Majorana particle that can self-annihilate.
Generically, this SuSy particle is referred to as the lightest stable particle (LSP). 

 The detection of gamma-rays with energies up to 10~TeV as seen with the
H.E.S.S. telescopes implies that the WIMP mass  has to be sufficienly large to
explain the observed spectrum. The surprisingly large mass inferred from the
observations appears at first glance to be inconsistent with cosmology and the
general conception that the mass of LSP in SuSy models should be close to the
electro-weak unification scale \cite{1981NuPhB.188..513W}. However, it has
been argued that a multi-TeV WIMP is in fact natural in Minimal Supergravity
models \cite{2000PhRvL..84.2322F}.  The upper bound on the possible mass of a
WIMP that decoupled thermally during the early phase of expansion is around
40~TeV \cite{1990PhRvL..64..615G} for a cosmological non-baryonic Dark Matter
density of $\Omega_{dm} h^2=0.1$.

The average  annihilation cross section $\langle \sigma v\rangle$ for SuSy
WIMPs is in the right range to explain today's Dark Matter density as
determined by the WMAP mission: $\Omega_{dm} h^2=0.116\pm0.020$
\cite{2003ApJS..148..175S}.  However, given that for the thermal decoupling
$\Omega_{dm} h^2 \propto \langle \sigma v\rangle^{-1}$ and that $\langle \sigma
v\rangle\propto m_\chi^{-1}$,
rather tight constraints on the mass of the WIMP $m_\chi<600$~GeV can be
derived \cite{2003PhLB..565..176E}. However, as already pointed out in
\cite{2003PhLB..565..176E} there are specific manifestations of SuSy models
(even in the constrained, universal models) that allow for larger WIMP masses
which still produce the right relic density:

\begin{enumerate}

\item For specific SuSy models in which the LSP is almost degenerate in mass
with other SuSy particle (e.g. $\tilde \tau$) coannihilation depletes the relic
density even for a small annihilation cross section.

\item For a non-thermal decoupling process, the relic density as it is today is
not depending on the cross section. 

\item Relaxing the universality of the Higgs multiplett masses at the GUT
scale: In this (and in the general SuSy) case, it is conceivable that a massive
(multi-TeV) WIMP with the right relic density may exist. Thess WIMPs are
generally almost pure Higgsino type Neutralinos with a large cross section.

\end{enumerate}

 In the first case, the cross section for annihilation remains small and 
therefore requires high relic density values in the Galactic center to explain the observed $\gamma$-ray
flux.  In the special case of a degeneracy between the LSP and  other SuSy
particles, careful tuning is required to achieve the right relic density. For a
massive LSP beyond a few TeV, even co-annihilation does not suffice to reduce
the relic density. A non-thermal scenario invoking Q-balls \cite{1985NuPhB.262..263C} 
 which replenish the Dark Matter content via evaporation at some time after the
decoupling has been studied in the literature \cite{1999NuPhB.538..321E,2002PhRvD..66h3501F}.
In this and other non-thermal scenarios the cross section for WIMPs could be rather high.

 The last case is more attractive but again requires tuning
of the Gaugino masses at the unification scale which might appear less
attractive from the point of view of a natural theory.

 Generally, a massive WIMP as suggested by the observations with a sufficiently
high cross section is not the obvious candidate in the framework of MSUGRA but
could be the LSP in slightly less general theories.  In this way, thermal
decoupling would produce the right relic density and the observed flux could be
consistent with a Dark Matter density which is not violating other
astrophysical bounds.

 Ultimately, a neutralino mass exceeding 40~TeV is excluded by the unitarity
bound in the framework of thermal decoupling \cite{1990PhRvL..64..615G}.

\section{WIMP density}
 Given the measured flux, it is possible (without assuming a specific halo
shape) to derive an upper limit on the product $\langle \rho_{dm}\rangle_{r<R_0}
\sqrt{\langle \sigma v\rangle}$ This is truly an upper limit, as can be seen
from the following argument:\\ Assuming that the true density distribution is
spherically symmetric and follows a general form \cite{1998APh.....9..137B}
\begin{eqnarray}
\rho(r)&=&\frac{1}{(r/a)^\gamma[1+(r/a)^\gamma]^{(\beta-\gamma)/\alpha}}\label{eqn:r}
\end{eqnarray} 
The halo density profiles predicted from N-body simulations of hierarchical structure formation 
follow a power law roughly between $\gamma=1$ \cite{1997ApJ...490..493N} and $\gamma=1.5$ \cite{1999ApJ...524L..19M}.
The annihilation flux observed from within the Galaxy is
proportional to the line of sight integral \begin{eqnarray} J(\psi)
&=& J_0\times\int_{l.o.s.} dl\, \rho^2\left(\sqrt{R^2-2lR\cos\psi+l^2}\right)
\end{eqnarray} averaged over the solid angle observed: \begin{eqnarray} \jpsiDO
&=& J_0\times \int d\Omega\, \int dl\,\rho^2 \label{eq:los} \end{eqnarray} with
the commonly used normalization for the Galactic Center: \begin{eqnarray} J_0
&=& \frac{1}{8.5\,\mathrm{kpc}}
\frac{1}{(0.3~\mathrm{GeV}\,\mathrm{cm}^{-3})^2}=4.2\times
10^{-22}\,\mathrm{GeV}^{-2}\,\mathrm{cm^5} \end{eqnarray} For reasons of
simplification and in order to calculate an upper limit, it is convenient to
use a truncated version of Eq.~\ref{eqn:r} which drops
to zero outside a spherical volume with radius $R_0$.  Obviously, the derived 
density to match an observed flux would be higher than in the case of a
continuous halo distribution.  
For a truncated density function with $R_0$ much smaller than the distance of the observer to the sun, the line of sight integral can be safely 
replaced by a volume integral and the observed differential gamma-ray flux is
given by: 
\begin{eqnarray} \frac{d\Phi}{dE} & =& \frac{1}{2}\frac{1}{4\pi
d^2}\,\frac{dN}{dE}\,\frac{\langle \sigma v\rangle}{m_\chi^2}\times\int
dV\,\rho^2 \nonumber \\ & \propto& \langle \rho^2 \rangle V
\label{eq:flux} \end{eqnarray}

Considering a truncated density function which follows a single power law up to 
a radius $R_0$ dropping to zero for larger radii: 
$\rho(r)=\rho_0\times(R_0/r)^\alpha$ with $\alpha=0\ldots1.5$. 

\begin{eqnarray}
 \langle \rho \rangle &=& \frac{\int dV\rho}{\int dV} \nonumber \\
                      &=&\rho_0\times \frac{3}{3-\alpha}
\end{eqnarray}

and for $\langle \rho^2\rangle$:
\begin{eqnarray}
 \langle \rho^2 \rangle &=& \frac{\int dV\rho^2}{\int dV} \nonumber \\
		         &=& \rho_0^2\times\frac{3}{3-2\alpha}
\end{eqnarray}
 which gives for the ratio $\langle \rho^2 \rangle/\langle \rho \rangle^2$:
\begin{eqnarray}
  \frac{\langle \rho^2\rangle}{\langle \rho \rangle^2} &=& \frac{(3-\alpha)^2}{3(3-2\alpha)}
\end{eqnarray}
This ratio is $>1$ for $\alpha>0$. For $\alpha=0$ the 
trivial case of a constant density and $\langle \rho^2\rangle=\langle \rho\rangle^2$ is
recovered. For $\alpha>0$ $\langle \rho^2\rangle>\langle \rho\rangle^2$ which 
implies for a given flux ($\propto
\langle \rho^2\rangle$, see Eq.~\ref{eq:flux}) a \textit{lower} average density than in
the case of a constant WIMP density\footnote{For the sake of brevity, this is only
shown for $0<\alpha<1.5$, but it holds true for any $\alpha>0$.}. Therefore,
calculating an average density with $\alpha=0$ from the observed flux sets an upper limit
on the actual average density for a realistic radial density distribution.

 For convenience, Eq.~\ref{eq:flux} can be rewritten with appropriate units:
\begin{eqnarray}
\frac{d\Phi}{dE} &=& 0.54\times10^{-12} (\mathrm{cm}^2\,\mathrm{s})^{-1}\,\frac{\langle \rho\rangle^2}{(5000\,\mathrm{GeV}\,c^{-2} \mathrm{cm}^{-3})^2}
\cdot\frac{\langle \sigma v\rangle}{3\times10^{-26}\mathrm{cm}^3\mathrm{s}^{-1}}\nonumber \\ 
\left(\frac{m_\chi}{\mathrm{TeV}}\right)^{-2}&\times&
\frac{(3-\alpha)^2}{3(3-2\alpha)} 
\times\left(\frac{d}{8.5~\mathrm{kpc}}\right)^2 
\times\left(\frac{R_0}{10~\mathrm{pc}}\right)^3 \frac{dN}{dE} \label{eqn:fluxexp}
\end{eqnarray}
 In the more general case of a radial density profile extending beyond $R_0$, the
 line-of-sight integral of Eq.~\ref{eq:los}  will naturally give larger
 observed fluxes ($\propto \jpsiDO$) than for a truncated density therefore relaxing
 the requirement of the average central density $\langle \rho\rangle$ even more. In 
 this sense, the approach of calculating an average, truncated density is truly 
 an upper limit on the average density as expected for a realistic, smooth Dark Matter
 halo distribution.

 The parameterization for the decay spectrum is of course depending on the
nature of the WIMP and cannot be given in general. For the case of 
a very massive WIMP considered here, the dominant annihilation channel is
into a $W^+W^-$, $Z^0Z^0$, and $q\bar q$ pairs. 
A parameterization of the form
\begin{eqnarray}
\frac{dN}{dE}&=&\frac{0.73}{m_\chi}\times\frac{e^{-7.8\,E/m_\chi}}{(E/m_\chi)^{1.5}+E_0}
\end{eqnarray}
is a reasonable description of the decay spectrum (with $E$ and $m_\chi$ in TeV, $E_0=2\cdot10^{-4}~\mathrm{TeV}$).

In order to calculate the upper limit on $\langle \rho\rangle$ it is assumed
that the entire observed flux is produced by annihilation of WIMPs.  The
H.E.S.S. data-points are taken from \cite{Nature} with the assumed emission
region being covered by the point-spread function of $\approx 5.8$'. The actual
limit on the source size is only 3'. However, this assumes a Gaussian surface
emission profile, whereas a volume source with constant emissivity is expected.
The upper limit on such an emission profile will be somewhat bigger than 3'
but smaller than the point spread function. The upper limit depends on the
assumed emission region as $\langle \rho\rangle\propto R_0^{-3/2}$. 

\begin{figure}
\centering
 \includegraphics[width=\linewidth]{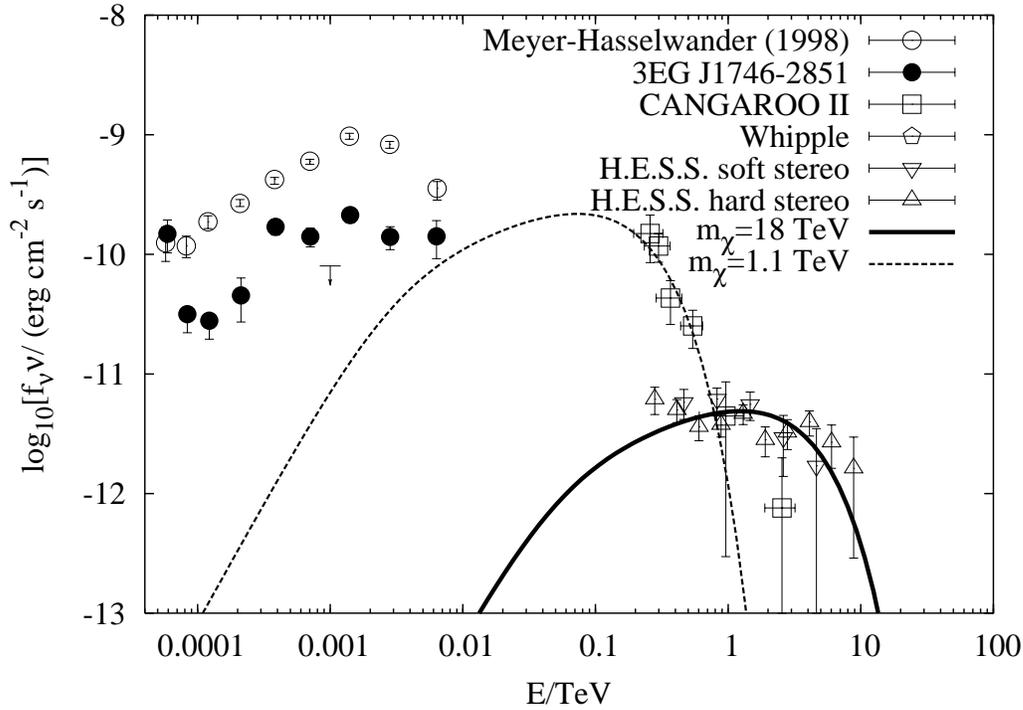}
\caption{A summary of data and \label{fig:data} best-fit models for WIMP annihilation from the Galactic center: H.E.S.S. (open triangles), CANGAROO (open boxes), EGRET (solid and open circles), 10m Whipple telescope of the VERITAS collaboration (solid diamond).}
\end{figure}

The H.E.S.S. data points (open triangles) are indicated in Fig.~\ref{fig:data} together with the CANGAROO (open boxes), VERITAS (open diamond), and different
EGRET flux values (open symbols from \cite{1998A&A...335..161M}, 
filled symbols from \cite{1999ApJS..123...79H}, and upper limit from \cite{2002astro.ph.10617H}).
 A best fit model of the H.E.S.S. 
data requires 
\begin{eqnarray*}
m_\chi&=&(18.9\pm3.7)~\mathrm{TeV} \\
\langle \rho\rangle_{r<10\mathrm{pc}} &=&(47900\pm4700)~\frac{\mathrm{GeV}}{\mathrm{c}^{2}\mathrm{cm}^{3}}
\times \left(\frac{\langle \sigma v\rangle}{3\times10^{-26}\,\mathrm{cm}^3\,\mathrm{s}^{-1}}\right)^{-1/2}\frac{\sqrt{3(3-2\alpha)}}{3-\alpha}\\
\\
 & =& (1261\pm124)~\frac{M_\odot}{\mathrm{pc}^{3}}
\times \left(\frac{\langle \sigma v\rangle}{3\times10^{-26}\,\mathrm{cm}^3\,\mathrm{s}^{-1}}\right)^{-1/2}\frac{\sqrt{3(3-2\alpha)}}{3-\alpha}
\end{eqnarray*} 
with $\chi^2_{red}(d.o.f.)=1.43(13)$.
 
\begin{figure}
\centering
\includegraphics[width=\linewidth]{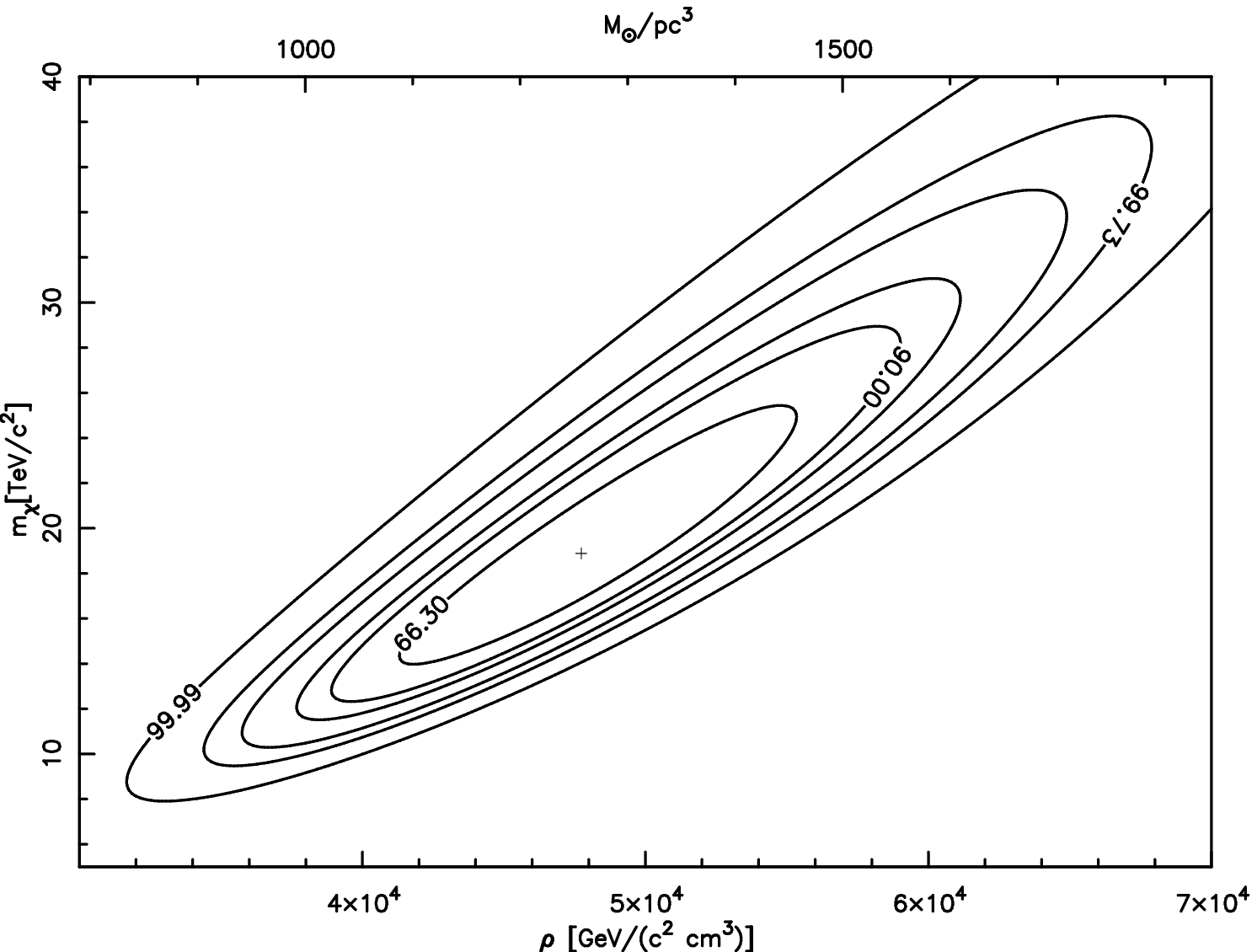}
\caption{\label{cont} For the $\chi^2$ fit of the model energy spectrum
to the data confidence regions are indicated for different confidence values.
The best-fit model is indicated with a small cross.}
\end{figure}

 The confidence regions given by the $\chi^2$ minimization for the
two parameters $\langle \rho\rangle$ and $m_\chi$ is given in Fig.~\ref{cont}. The best fit position is indicated by a small cross.

For the sake of completeness, a similar procedure can be applied to the CANGAROO data giving the following values:
\begin{eqnarray*}
m_\chi&=&(1.1\pm0.3)~\mathrm{TeV}\\
\langle \rho\rangle_{r<47\mathrm{pc}} &=&(7840\pm1170)~\frac{\mathrm{GeV}}{c^{2}\mathrm{cm}^{3}}
\times \left(\frac{\langle \sigma v\rangle}{3\times10^{-26}\,\mathrm{cm}^3\,\mathrm{s}^{-1}}\right)^{-1/2} \frac{\sqrt{3(3-2\alpha)}}{3-\alpha}\\
\\
  &=&(206\pm 31)~\frac{M_\odot}{\mathrm{pc}^3} 
\times \left(\frac{\langle \sigma v\rangle}{3\times10^{-26}\,\mathrm{cm}^3\,\mathrm{s}^{-1}}\right)^{-1/2} \frac{\sqrt{3(3-2\alpha)}}{3-\alpha}
  \end{eqnarray*}
with $\chi^2_{red}(d.o.f.)=0.81(4)$. The value given for the density is
higher than calculated in \cite{2004ApJ...606L.115T} because of the
factor of $1/2$ in Eqn.~5 that has been neglected in the past 
\cite{2004astro.ph..6204G}. As pointed out in the Introduction, the CANGAROO-II
result is not consistent with the H.E.S.S. observations hinting 
at either a mismatch in the cross-calibration of the two instruments 
or a variable object which might coincide with one of the more conventional gamma-ray 
sources close to the Galactic center.

The required average density depends on the cuspiness of the halo profile and 
scales according to Eqn.~\ref{eqn:fluxexp} with $\sqrt{3(3-2\alpha}/(3-\alpha)$ (for $\alpha<1.5$) and
is reduced by a factor of 2 with respect to a 
constant density halo distribution when $\alpha\approx1.4$.

\section{Spatial WIMP distribution}

 Given the excellent angular resolution of the H.E.S.S. telescope arrays of 
 5.8' for individual events  \cite{Nature}, 
 spatial extension of the 
 source down to this level can be probed. The signal appears to be point-like (i.e.
 of spatial extension smaller than the point spread function) with an
 upper limit of 3' for a source with a Gaussian surface brightness profile. 

 The expected source extension from Dark Matter annihilation depends crucially 
 on the shape of the radial density profile. For a quantitative study, a 
 simplification of the radial density profile  given in Eqn.~\ref{eqn:r} 
 is applied:
 \begin{eqnarray}
\rho(r) &=& \rho_0\times\left( \frac{a}{r}\right)^\alpha  (r<a=8.5\mathrm{kpc})\\
        &=& 0 \,\,\,(r>a)
\end{eqnarray}
Since the observations cover only the inner 0.5$^\circ$ of the Galactic center (equivalent to
75~pc), the actual value of $a$ is not crucial.

\begin{figure}
\centering
\includegraphics[width=\linewidth]{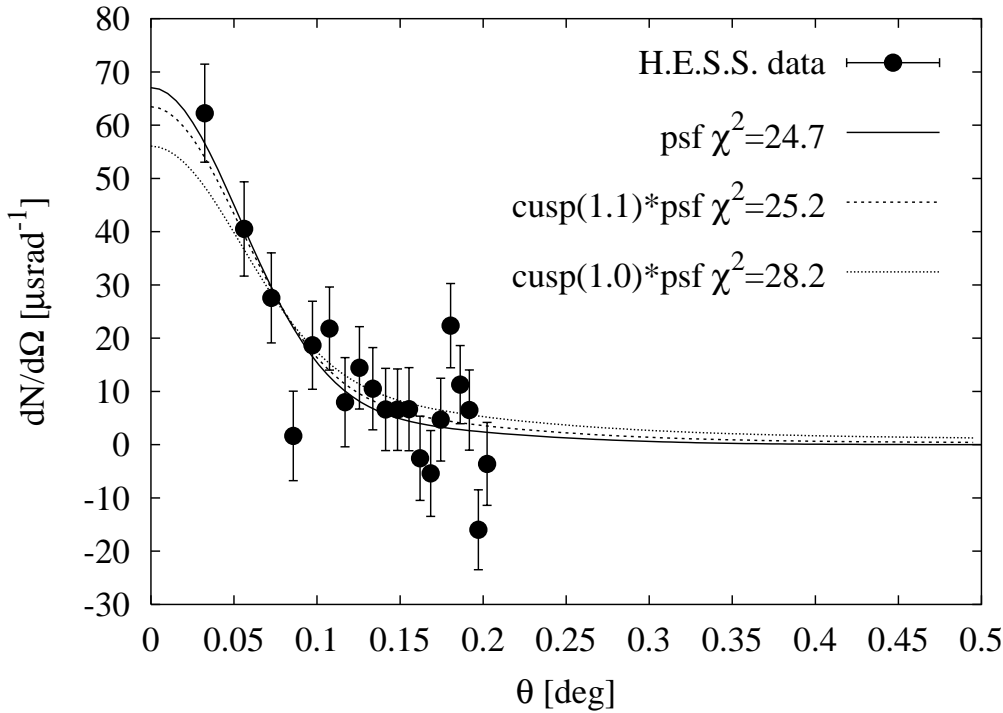}
\caption{For different values of the slope of the density profile, the
expected  observed angular distribution (after folding with the point spread
function) allows to separate the case of $\alpha<1.0$ and $\alpha>1.0$. From
\label{fig:ang}
the good agreement of data with a point source, a steep (cuspy) halo is
required.}
\end{figure}

 The observed surface brightness is calculated by folding $J(\psi)$ (Eqn. 2) with the 
 average point spread function of H.E.S.S. for the observations  approximated
by a double Gaussian 
\begin{eqnarray}
f_{psf}(\psi) &=& f_0 \cdot (\exp(-0.5\psi^2/\sigma_1^2)+1/8.7\cdot\exp(-0.5\psi^2/\sigma_2^2))
\end{eqnarray}
 with $\sigma_1=0.052^\circ$, $\sigma_2=0.136^\circ$ (J. Hinton, priv. communication), and
$f_0$ as an arbitrary normalization:
 \begin{eqnarray}
   p(\psi) &=& p_0\times\int\limits_{-\infty}^{\infty}
    d\psi' \cdot f_{psf}(\psi-\psi') J(\psi')
 \end{eqnarray}
  with $p_0$ chosen such that $\int d\psi\,p(\psi)=1$. The resulting 
  angular distribution is shown in Fig.~\ref{fig:ang} for different values
  of $\alpha$.  Using the $\chi^2$ values for various $\alpha$, a 90~\%
c.l. (one sided) for $\alpha>1$ is calculated which is suggestive of a 
halo with a halo steeper than predicted by Navarro, Frenk, and White \cite{1997ApJ...490..493N}.

\section{Discussion}
TeV gamma-ray emission from the Galactic center has been observed by different
different ground-based instruments. The CANGAROO group claims 
a signal at a
significance close to 10~$\sigma$ and the VERITAS group sees a 
indications for a signal observed 
in archival data.  Finally, the H.E.S.S. system reported 
an overall excess of 11.8~$\sigma$ from the Galactic center region from observations during the build-up phase.


However, taking the H.E.S.S. data alone, no indication for variability is seen
and moreover, the source location is consistent with the position of \sgra 
($\Delta \theta= (18\pm42)$'' with a systematic error of $\approx 20$'') and an
upper limit for a Gaussian surface density source of $3$'.
 
 Taking these results, a hypothetical Dark Matter annihilation scenario is
discussed.  The observed energy spectrum is reasonably well described by the
continuum emission from SuSy WIMPS (neutralinos) with a mass of 18~TeV (lower
limit of \ml) putting it far out of reach for future accelerator experiments.

 The required average Dark Matter density is $(1092\pm107)\cdot
M_\odot\,\mathrm{pc}^{-3}$ (for $\langle \sigma v\rangle = 3\cdot
10^{-26}\,\mathrm{cm}^3\mathrm{s}^{-1}$ and $\alpha=1$) for the inner 10~pc of
the Galaxy with a cuspy profile $\rho\propto r^{-\alpha}$ with $\alpha>1$ in
order to fit the observed angular distribution.  The upper limit on nonbaryonic
Dark Matter density is close to the stellar (baryonic) mass density which is
$\rho_*(r)\approx (1900\pm700)\times (r/10\mathrm{pc})^{-1.4\pm0.1}
~M_\odot\,\mathrm{pc}^{-3} $\cite{genzel}. The total inferred mass
$M(r<10~\mathrm{pc})=(3.1\pm0.3) \times 10^6~M_\odot$ is well below the stellar
mass of $(1.5\pm0.6)\times10^7~M_\odot$ in the inner 10~pc. The rather high
central values of the density is consistent with the formation of a
\textit{mini-cusp} as a consequence of interaction with the central dense
stellar environment \cite{primack,2004PhRvL..92t1304M} and baryonic compression
\cite{prada}. 

The synchrotron emission emitted by the electrons/positrons injected through
the annihilation process constrains the maximum slope of the radial density distribution \cite{gondolo,aloisio} which rules out a \textit{spike} close
to the black hole \cite{spike}. The constraint is further relaxed if fast
diffusion or convection removes the electrons/positrons from the vicinity 
of the black hole.


The angular resolution in gamma-rays achieved with H.E.S.S. allows to probe
the actual shape of the Dark Matter profile
for the Galactic center. With more data taken with the full telescope array
operational, a further improvement on the constraint on the cuspiness is
expected and test the different possibilities for the origin of the signal.

At this point, the considered annihilation scenario of WIMPS as the origin of
the observed flux of gamma-rays is not ruled out. The other detectable channels
of annihilation i.e. excess of positrons and anti-protons in Cosmic-rays are
not affected by a mini-cusp scenario as outlined above. The large mass of the
WIMP may also produce a signal for neutrino-telescopes but the sensitivity of
the current instruments is not sufficient to detect neutrinos from the Galactic
center \cite{2004astro.ph..3322B}. In principle, bounds from big-bang
nucleosynthesis could be of importance but as outlined in \cite{jedamzik}, a
very massive WIMP would not have a strong impact on abundances of light
elements.  Future observations of variability in the gamma-ray flux or a possible
extension would be of great importance for the interpretation of the signal.

The author is grateful for discussions with F. Aharonian,
J. Buckley, A. Jacholkowska, J. Primack, P. Ullio,  H.J. V\"olk, and members of the H.E.S.S. 
working group on Dark Matter.

\end{document}